\documentclass[aps,prl,twocolumn,superscriptaddress]{revtex4-1}
\usepackage{graphicx}
\usepackage{color}
\usepackage{amsfonts}
\usepackage{isomath}
\usepackage{amsmath}
\usepackage{amsthm}
\usepackage{amsfonts}
\usepackage{braket}

\usepackage{yhmath}

\def\kbar{$\bar{\mathrm{K}}$}

\def\WS2{WS$_2$}
\def\MoS2{MoS$_2$}{

\begin{document}

\title{Momentum-resolved view of highly tunable many-body effects in a graphene/hBN field-effect device}

\author{Ryan Muzzio$^{\star}$}
\affiliation{Department of Physics, Carnegie Mellon University, Pittsburgh, Pennsylvania 15213, USA}
\author{Alfred J. H. Jones$^{\star}$}
\author{Davide Curcio}
\author{Deepnarayan Biswas}
\author{Jill A. Miwa}
\author{Philip~Hofmann}
\affiliation{Department of Physics and Astronomy, Aarhus University, 8000 Aarhus C, Denmark}
\author{Kenji~Watanabe}
\author{Takashi~Taniguchi}
\affiliation{National Institute for Materials Science, 1-1 Namiki, Tsukuba 305-0044, Japan}
\author{Simranjeet Singh}
\affiliation{Department of Physics, Carnegie Mellon University, Pittsburgh, Pennsylvania 15213, USA}
\author{Chris Jozwiak}
\affiliation{Advanced Light Source, E. O. Lawrence Berkeley National Laboratory, Berkeley, California 94720, USA}
\author{Eli Rotenberg}
\affiliation{Advanced Light Source, E. O. Lawrence Berkeley National Laboratory, Berkeley, California 94720, USA}
\author{Aaron Bostwick}
\affiliation{Advanced Light Source, E. O. Lawrence Berkeley National Laboratory, Berkeley, California 94720, USA}
\author{Roland J. Koch}
\affiliation{Advanced Light Source, E. O. Lawrence Berkeley National Laboratory, Berkeley, California 94720, USA}
\author{S{\o}ren Ulstrup}
\email{Address correspondence to: ulstrup@phys.au.dk}
\affiliation{Department of Physics and Astronomy, Aarhus University, 8000 Aarhus C, Denmark}
\author{Jyoti Katoch}
\email{Address correspondence to: jkatoch@andrew.cmu.edu}
\affiliation{Department of Physics, Carnegie Mellon University, Pittsburgh, Pennsylvania 15213, USA \\ $^{\star}$ These authors contributed equally}

\begin{abstract}
Integrating the carrier tunability of a functional two-dimensional material electronic device with a direct probe of energy- and momentum-resolved electronic excitations is essential to gain insights on how many-body interactions are influenced during device operation. Here, we use micro-focused angle-resolved photoemission in order to analyze many-body interactions in back-gated graphene supported on hexagonal boron nitride. By extracting the doping-dependent quasiparticle dispersion and self-energy, we observe how these interactions renormalize the Dirac cone and impact the electron mobility of our device. Our results are not only limited to a finite energy range around the Fermi level, as in electron transport measurements, but describe interactions on a much wider energy scale, extending beyond the regime of hot carrier excitations.
\end{abstract}

\maketitle

The realization of a two-dimensional (2D) gas of Dirac electrons with a density that can be tuned over several orders of magnitude has triggered numerous tantalizing discoveries of unconventional electronic behavior in graphene \cite{Novoselov:2004,Zhang:2005,Novoselov:2005}, including a departure from normal Fermi liquid theory \cite{sarmamanybody2007,Elias:2011} and the appearance of a fractional quantum Hall effect \cite{Xu:2009,Bolotin:2009}. Accessing doping-dependent many-body interactions in graphene is routinely achieved in a noninvasive manner through electrostatic doping using a back-gated device configuration in transport \cite{Novoselov:2004,Zhang:2005,Novoselov:2005,Mayorov:2012,Efetov:2010} and scanning tunneling spectroscopy experiments \cite{Zhang:2008}. It would be highly desirable to use angle-resolved photoemission (ARPES) as a complementary tool because it provides the full energy and momentum dependent spectral function of the occupied states, thereby unveiling how the Dirac cone renormalizes in the presence of quasiparticle scattering \cite{Bostwick:2007,hwangquasiparticle2008,Basov:2014}.

Extracting many-body effects in graphene for different carrier densities using ARPES is commonly achieved by depositing alkali metal atoms, which act as electron donors. This approach has been remarkably successful for measuring electron-hole and electron-plasmon excitations \cite{Bostwick:2007,Bostwick:2010}, but it has the major drawback of being irreversible and thus difficult to use for fine-tuning the carrier density. Adsorbed adatoms are also known to act as impurities, causing scattering, which results in an increase in measured linewidths that can be difficult to deconvolve from intrinsic interactions \cite{Chen:2008,Siegel:2013,Ulstrup:2014}. Alternatively, one may change the doping in graphene by replacing the supporting substrate \cite{Siegel:2011,Ulstrup:2016}, but this inadvertently changes the background dielectric screening of charge carriers. As a result it becomes difficult to unambiguously correlate many of the phenomena observed in ARPES with standard device measurements of graphene, when using these irreversible doping methods to change the carrier density. It is therefore of utmost importance to merge ARPES measurements with in situ electric-field doping in gated 2D material based devices.

\begin{figure} 
\begin{center}
\includegraphics[width=0.49\textwidth]{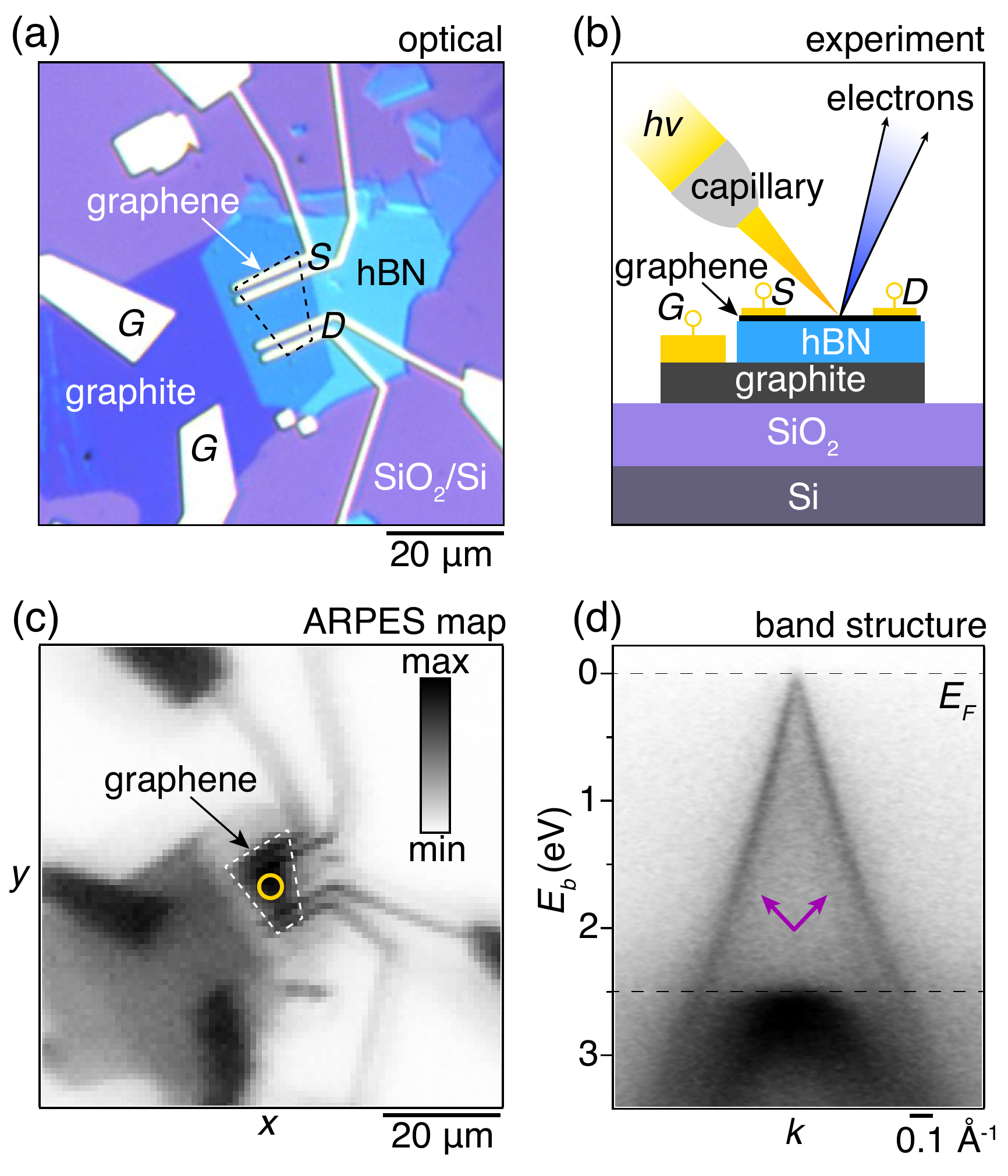}
\caption{(a) Optical micrograph of device. (b) Sketch of the experimental setup integrating micro-scale photoemission with source ($S$), drain ($D$) and gate ($G$) contacts fabricated on a graphene/hBN/graphite stack placed on a SiO$_2$/Si wafer. (c) Map of  $(x,y)$-dependent photoemission intensity corresponding to the region shown in (a). The graphene flake is demarcated by dashed lines and an arrow in (a) and (c). (d) Dirac dispersion collected from the graphene flake in the spot marked by a yellow circle in (c). The purple arrows point to faint minibands originating from the graphene/hBN interface. The map in (c) is composed from the integrated intensity enclosed between the dashed lines in (d).}
\label{fig:1}
\end{center}
\end{figure}

The mesoscopic sizes and intrinsic inhomogeneities of such devices have posed the biggest challenges precluding conventional ARPES studies. These issues can be circumvented by using a microscopically focused beam of photons as demonstrated in recent micro-focused angle-resolved photoemission (microARPES) experiments performed on 2D material based heterostructures and devices \cite{katoch2018,Joucken:2019,Nguyen:2019,ulstrup2019direct}. We apply this approach here to investigate the Coulomb interaction in graphene on hexagonal boron nitride (graphene/hBN) at a relatively small interlayer twist angle of 2.0$^{\circ}$. This stack has been integrated in a device architecture with a graphite back-gate, as shown in the optical micrograph in Fig. \ref{fig:1}(a) \cite{SMAT}. Electrical doping of the device is achieved by grounding the graphene flake and applying a constant voltage to the graphite back gate. Maximum hole ($p$) and electron ($n$) type dopings were determined by the onset of a leakage current from the back gate through the hBN insulating layer. The sample, consisting of device and wire-bonds, was given a mild anneal to 150~$^{\circ}$C for 3~hours on a hot plate inside a water- and oxygen-free glovebox connected to the same ultra-high vacuum environment as the microARPES system.

The measurements were carried out at the MAESTRO facility at the Advanced Light Source, Lawrence Berkeley National Laboratory. Using an achromatic focusing capillary, which simultaneously provides a high photon flux and a  $(1.8 \pm 0.3)$ $\mu$m beam-spot \cite{ulstrup2019direct}, we are able to collect high quality microARPES spectra that allow for a detailed analysis of many-body effects in graphene.  The photon energy was set at 90 eV unless otherwise noted, and the optimum achieved energy- and $k$-resolution were 40~meV and 0.01~\AA$^{-1}$, respectively. The measurements were carried out at room temperature. Our experimental configuration, which is sketched in Fig. \ref{fig:1}(b), fully realizes \textit{in operando} microARPES.  

Our device is initially mapped by scanning the photon beam from the capillary over the same area as seen in the optical micrograph in Fig. \ref{fig:1}(a). The resulting $(x,y)$-dependent photoemission intensity is presented in Fig. \ref{fig:1}(c). A corresponding snapshot of the $E(k)$-dispersion from the region marked by a circle in the map is shown in Fig. \ref{fig:1}(d). One immediately notices a sharp Dirac cone due to the graphene flake in this region, as well as the onset of an intense band at a binding energy of 2.5~eV. The latter is consistent with the dispersion around the hBN valence band maximum \cite{Koch:2018}. The map in Fig. \ref{fig:1}(c) has been composed from the $k$- and $E$-integrated intensity in the region enclosed by the dashed lines in Fig. \ref{fig:1}(d), making the graphene flake clearly distinguishable between source and drain electrodes due to the presence of the graphene Dirac cone. We refer to the supplemental material for further characterization of the features in the map \cite{SMAT}. The arrows in Fig. \ref{fig:1}(d) point to faint Dirac cone replicas that are consistent with the mini Brillouin zone of the graphene/hBN superlattice defined by the twist angle of 2.0$^{\circ}$ \cite{Wang:2016b,ulstrup2019direct,SMAT}.  

\begin{figure*} 
\begin{center}
\includegraphics[width=0.9\textwidth]{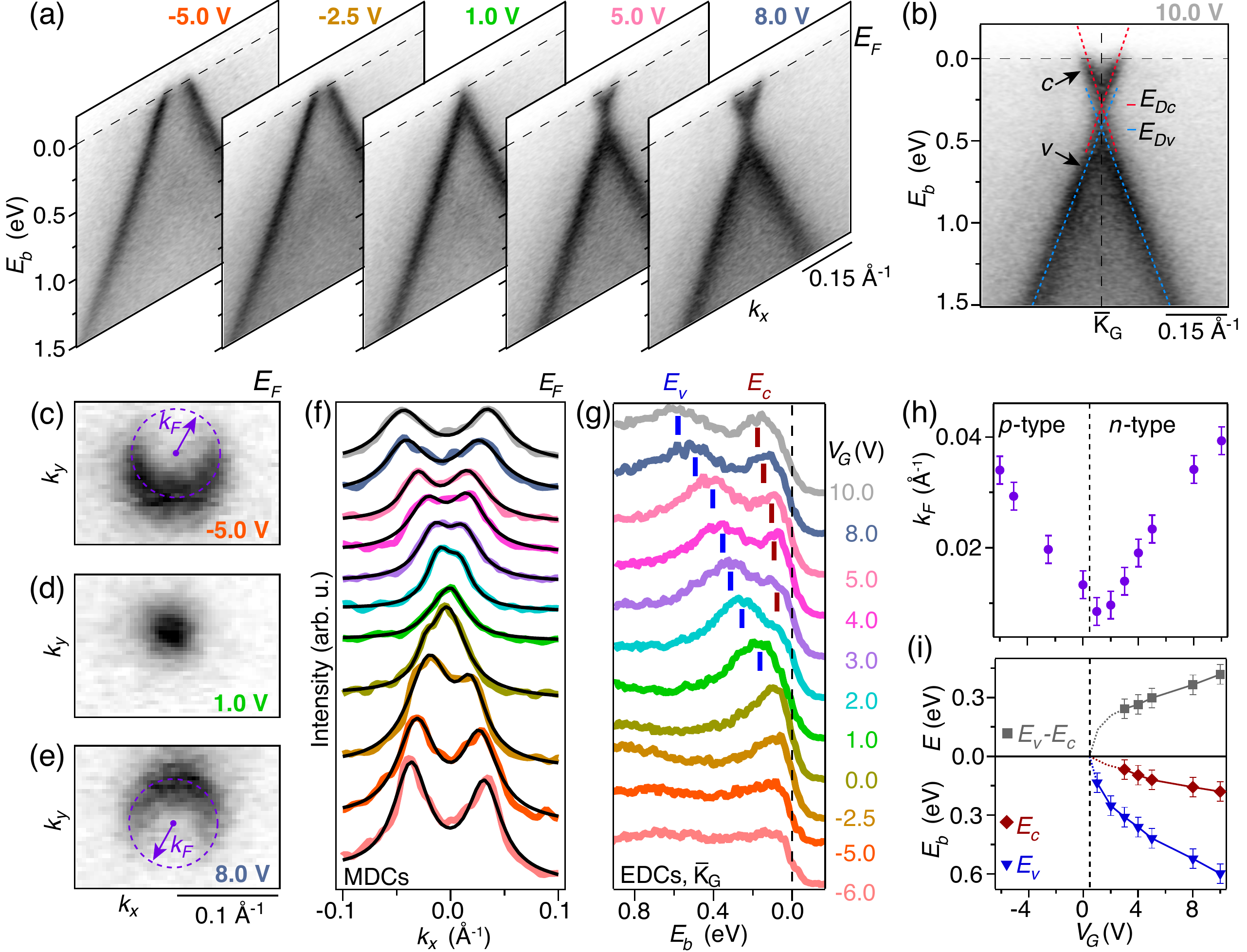}
\caption{(a) Snapshots of Dirac dispersion at the given gate voltages. (b) Dirac cone at the highest achieved gate voltage for $n$-type doping. The branches below (above) the Dirac crossing are labeled $v$ ($c$). The dashed curves represent linearly extrapolated bands determined from momentum distribution curve (MDC) fits to the $v$-branches (blue dashed bands) and $c$-branches (red dashed bands) \cite{SMAT}. The $v$ ($c$) branches cross at $E_{Dv}$ ($E_{Dc}$). (c)-(e) Constant energy contours at $E_F$ for (c) $p$-type doping, (d) near charge neutrality and (e) $n$-type doping. The purple dashed circle outlines the circular Fermi surface with radius $k_F$ indicated by a purple arrow. (f) MDCs extracted at $E_F$ (thick colored curves) with Lorentzian fits (black thin curves).(g) Energy distribution curves (EDCs) at \kbar$_{\mathrm{G}}$ with tick marks indicating simple estimates of the $v$-crossing $E_v$ (blue tick) and the $c$-crossing $E_c$ (red tick). Each curve in (f)-(g) corresponds to a different value of $V_G$ as noted on the right of panel (g). (h)-(i) Gate voltage dependence of (h) $k_F$ determined from the MDC peak positions in (f), and (i) the EDC peak positions in (g) along with their separation ($E_v - E_c$).}
\label{fig:2}
\end{center}
\end{figure*}  
 
Detailed measurements of the $E(k)$-dispersion of the Dirac cone as a function of gate voltage ($V_G$) in our device are presented in Fig. \ref{fig:2}. The series of snapshots around \kbar$_{\mathrm{G}}$ in panels~(a)-(b) and in the Supplementary Video demonstrate excellent control of both $p$- and $n$-type fillings of the Dirac cone, thereby giving access to the valence ($v$) and conduction ($c$) bands defined in panel~(b). Each snapshot has been obtained from the full $E(k_x,k_y)$-dispersion around the Dirac cone, which simultaneously provides the circular Fermi surface, as shown for a subset of gate voltages in panels~(c)-(e). Recently, similar \textit{in situ} electrostatic doping experiments on graphene have displayed an energy gap at the Dirac crossing for $n$-type doping \cite{Nguyen:2019}. Such a gap is likely to result from a slight misalignment of the detector scattering plane from the crystal axes, or stray fields affecting the electron trajectory, and is avoided in our measurements by careful alignment of the detector at each gate voltage step.

The radius $k_F$ that determines the size of the Fermi surface (see panels~(c) and (e)) is found by extracting the $k$-separation between the peak positions of the two linear branches at $E_F$ for each gate voltage. The peak positions are obtained using double Lorentzian fits of momentum distribution curves (MDCs) as shown in panel~(f). The $V_G$-dependence of the crossings between $v$- and $c$-branches is monitored through energy distribution curve (EDC) cuts at \kbar$_{\mathrm{G}}$ as presented in panel~(g). The crossing of the $v$-branches is detected via a peak (see blue tick labeled $E_v$) that is fully below $E_F$ in our spectra for $V_G \geq 1.0$~V. We find that the crossing of the $c$-branches (see red tick labeled $E_c$) is separated in binding energy from the $v$-crossing and that its corresponding peak is fully below $E_F$ for $V_G \geq 3.0$~V. The full $V_G$-dependence of $k_F$ and of the separation ($E_v - E_c$) determined from the MDC and EDC analyses are shown in panels~(h)-(i). The doping dependence of $E_v - E_c$ signals the presence of many-body interactions, which are contained in the photohole spectral function, $\mathcal{A}(E,k)~=~\pi^{-1}|\mathrm{\Sigma}^{\prime\prime}(E)|/\left[(E-\hbar v^{\ast}(k)k)^2 +|\mathrm{\Sigma}^{\prime\prime}(E)|^2\right]$, that ARPES measures \cite{Bostwick:2007}. Here, $E(k) = \hbar v^{\ast}(k)k$ is the quasiparticle dispersion around the Dirac point with the $k$-dependent slope $v^{\ast}(k)$ \cite{DasSarma:2013}, and $\mathrm{\Sigma}^{\prime\prime}$ is the imaginary part of the self-energy that is proportional to the quasiparticle scattering rate. It is related to the full-width at half maximum ($\mathrm{\Delta} k_{\mathrm{FW}}$) of the MDC fits as $\mathrm{\Sigma}^{\prime\prime} = \hbar v^{\ast}\mathrm{\Delta} k_{\mathrm{FW}}/2$. 

\begin{figure*} 
\begin{center}
\includegraphics[width=0.99\textwidth]{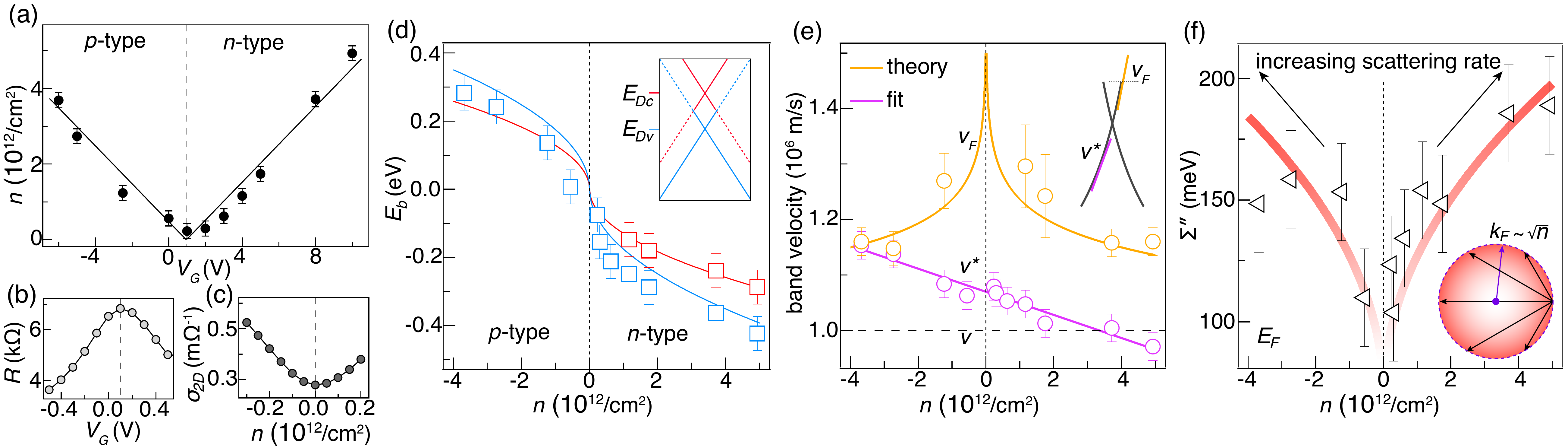}
\caption{(a) Carrier density determined as $k_F^2/\pi$ using the data in Fig. \ref{fig:2}(h). The curve is a fit to the expected linear dependence on $V_G$ \cite{Novoselov:2004}. (b) Resistance $R$ as a function of gate voltage obtained before exposure to the synchrotron beam \cite{SMAT}. (c) Sheet conductivity $\sigma_{2D}$ as a function of carrier density based on the fitted dependence in (a). (d) Binding energy positions of $E_{Dv}$ and $E_{Dc}$ (see definitions of these energies in the insert) determined by extrapolating the linear dispersion from below (blue squares) and from above (red squares) the Dirac point region, respectively. The curves are fits to $\sqrt{n}$-dependent functions. (e) Band velocities $v_F$ and $v^{\ast}$ measured at the Fermi level (orange circles) and 300~meV below $E_{Dv}$ (purple circles), respectively. The diagram illustrates the extraction of $v_F$ and $v^{\ast}$ from the slopes of the dispersion. The purple line is a fit to a linear dependence to guide the eye, and the orange curve is the analytic dependence of $v_F$ on doping for an effective Coulomb coupling constant given by $\alpha = 0.5$. The non-interacting velocity $v$ is indicated by a horizontal dashed line. (f) Imaginary part of the self-energy determined from the linewidth of Lorentzian fits to MDCs at the Fermi level. The curve is a fit to a $\sqrt{n}$-dependence, which follows from scattering processes that scale with the size of the Fermi surface $k_F$ as illustrated in the insert.}
\label{fig:3}
\end{center}
\end{figure*}

The many-body effects are investigated as a function of carrier density ($n$) using $n=k_F^2/\pi$ \cite{Efetov:2010}. Fitting $n$ to the expected linear dependence on $V_G$ demonstrated in Fig. \ref{fig:3}(a) and given by $n = C_a (V_G - V_0)/e$ \cite{Novoselov:2004}, where $e$ is the elemental charge, leads to estimates of the capacitance per area ($C_a = (80 \pm 5)$~nF/cm$^2$) and the shift of the charge neutrality point ($V_0 = (1.0 \pm 0.2)$~V). Combining this dependence with the resistance ($R$) of the device measured \textit{in situ} as seen in Fig. \ref{fig:3}(b) \cite{SMAT} provides the doping-dependent sheet conductivity $\sigma_{2D}(n)$ shown in Fig. \ref{fig:3}(c). This leads to the room temperature mobility $\mu = (6400 \pm 500)$~cm$^2$/Vs, which appears reasonable when comparing with other graphene/hBN devices \cite{Dean:2010}. We address the scattering processes at $E_F$ that affect $\mu$ towards the end of this letter. 

By combining the MDC fits at $E_F$ in Fig. \ref{fig:2}(f) with a more detailed analysis over a binding energy range of 1~eV measured from the Fermi level \cite{SMAT}, we extract the doping-dependent many-body interactions in our device. This analysis shows that it is reasonable to assume a linear dependence of the $v$ ($c$) band below (above) the Dirac crossing. We therefore linearly extrapolate the fitted dispersion as demonstrated in  Fig. \ref{fig:2}(b) to obtain a more precise estimate of the binding energy of the Dirac crossings than the $E_v$ and $E_c$ values picked out in Fig. \ref{fig:2}(g). The results are presented in Fig. \ref{fig:3}(d) for the $v$-crossing ($E_{Dv}$) for all dopings and the $c$-crossing ($E_{Dc}$) for the four most strongly $n$-type doped dispersions where the branches are sufficiently separated in $k$ to permit a reliable fit. The separation of the Dirac crossing is seen in Fig. \ref{fig:2}(i) and Fig. \ref{fig:3}(d) to vanish at charge neutrality, while it increases towards higher doping. This is a clear indication for electron-plasmon excitations \cite{Bostwick:2010}. We therefore use the one-particle Green's function calculations describing these excitations in Ref. \citenum{Walter:2011c} to convert the separation into an estimate for the effective Coulomb coupling constant $\alpha = e^2/4\pi\epsilon_0\epsilon\hbar v$, where $v=10^6$~m/s is the bare band velocity, $\epsilon_0$ is the vacuum permittivity and $\epsilon$ is the background screening constant. We obtain $\alpha \approx 0.5$ for our graphene/hBN device. In completely unscreened suspended graphene one would find $\alpha = 2.2$ \cite{Elias:2011,DasSarma:2013}. The smaller value of $\alpha$ in our sample is expected due to the underlying hBN, however the electron-electron interaction can still be rather substantial as observed in the spectral function of hydrogen intercalated graphene on silicon carbide where $\alpha$ has a similar value as we find here on hBN \cite{Bostwick:2010,Walter:2011c}.  

We investigate how the interaction strength in our device affects the doping-dependent shape of the Dirac cone by extracting the Fermi velocity ($v_F$) and the band velocity below the Dirac crossing. The values of $v_F$ for the dopings where we could perform a reliable fit are presented in Fig. \ref{fig:3}(e) together with the analytic result for the renormalized Fermi velocity $v_F/v = 1 - (\alpha/\pi)\left(\ln\alpha + 5/3\right) +(\alpha / 4)\ln\left(1/ak_F\right)$ \cite{DasSarma:2013}, where $a = 2.46$~\AA~is the graphene lattice parameter. The theoretical result is shown via the orange curve for $\alpha = 0.5$ and exhibits a trend that is consistent with our data. Note that $v_F$ would be constant and equal to $v$ as shown by a horizontal dashed line in Fig. \ref{fig:3}(e) in the fully screened limit given by $\alpha \rightarrow 0$. This is clearly not the case here where we instead find a sharpening effect of the cone towards charge neutrality, alluding to the situation in suspended graphene \cite{Elias:2011}. A substantially different behavior is found for the band velocity $v^{\ast}$ determined over an $(E,k)$-range centered 300 meV below $E_{Dv}$ for all dopings in order to avoid confusing changes of the slope with the bare band dispersion \cite{parkvan2008}. We find a continuous decrease of $v^{\ast}$ as the Dirac cone shifts to higher binding energies with increasing $n$-type doping, as shown in Fig. \ref{fig:3}(e), which is caused by the growing number of electron-hole pair scattering processes towards higher binding energies \cite{Bostwick:2007,sarmamanybody2007,Siegel:2011}. Our analysis affirms that the quasiparticle velocity in gated graphene on hBN is significantly doping-dependent due to the Coulomb interaction \cite{sarmamanybody2007}.  

In Fig. \ref{fig:3}(f) we examine the doping dependence of $\mathrm{\Sigma}^{\prime\prime}$ at $E_F$ determined from the MDC linewidths in Fig. \ref{fig:2}(f). The behavior of $\mathrm{\Sigma}^{\prime\prime}$ is characterized by a suppression around charge neutrality and a near-symmetric increase with both $p$- and $n$-type doping, which appears to follow a $\sqrt{n}$-dependence with a constant offset of $\sim$70~meV. From this we can exclude doping-dependent long-range charge impurity scattering, because this leads to a $1/\sqrt{n}$-dependence of $\mathrm{\Sigma}^{\prime\prime}$ as found in alkali metal doped graphene on hBN \cite{Siegel:2013}. The constant offset and $\sqrt{n}$-dependence are consistent with short-range electron-defect scattering and electron-phonon interactions \cite{Calandra:2007}, which scale with the radius of the Fermi surface $k_F \propto \sqrt{n}$ \cite{Efetov:2010}, as illustrated in the insert in Fig. \ref{fig:3}(f). These scattering processes at $E_F$ are ultimately responsible for the reduced mobility of our device.

We have simultaneously measured the transport properties and the doping dependence of the quasiparticle spectral function in graphene on hBN by noninvasively changing the carrier concentration with an electric field, providing access to many-body interactions for a wide range of energies and momenta. Combining the measurement of a global property such as electron mobility with $E$-, $k$- and spatially-resolved electronic excitations is a powerful approach that will be transformative for correlating fundamental interactions with different types of electronic behaviors observed in transport studies of quantum materials, including complex properties such as charge ordering and high temperature superconductivity. The intrinsic doping dependence of the spectral function holds the key to fully understanding the physics of these phenomena.  

\begin{acknowledgments}
J.K. is supported by the U.S. Department office of Science, Office of Basic Sciences, of the U.S. Department of Energy under Award Number DE-SC0020323. S. U. acknowledges financial support from VILLUM FONDEN under the Young Investigator Program (Grant No. 15375). The authors also acknowledge the Villum Centre of Excellence for Dirac Materials (Grant No. 11744). The Advanced Light Source is supported by the Director, Office of Science, Office of Basic Energy Sciences, of the U.S. Department of Energy under Contract No. DE-AC02-05CH11231. K.W. and T.T. acknowledge support from the EMEXT Element Strategy
Initiative to Form Core Research Center, Grant Number JPMXP0112101001 and the CREST(JPMJCR15F3), JST.
\end{acknowledgments}

\end{document}